\documentclass[12pt]{article}

\newcommand{\beq}{\begin{equation}}
\newcommand{\eeq}{\end{equation}}
\newcommand{\bea}{\begin{eqnarray}} 
\newcommand{\eea}{\end{eqnarray}}
\newcommand{\vs}[1]{\vspace{#1 mm}}

\renewcommand{\c}{\gamma}

\newcommand{\la}{\lambda}
\newcommand{\da}{\dot{a}}
\newcommand{\db}{\dot{b}}

\newcommand{\nn}{\nonumber\\ }

\def \da{{\dot a}}
\def \db{{\dot b}}

\def \thetat {\tilde{\theta}}

\begin{document}

\topmargin30pt
\oddsidemargin 0mm
\renewcommand{\thefootnote}{\fnsymbol{footnote}}
\begin{titlepage}

\setcounter{page}{0}

\vs{10}
\begin{center}

{\Large\bf  PP-wave String Interactions from String Bit Model}

\vs{15}

{\large Jian-Ge Zhou}\footnote{jiange.zhou@uleth.ca}

\vs{10}
{\em Physics Department, University of Lethbridge,
Lethbridge, Alberta, Canada T1K 3M4} \\
\end{center}

\vs{15}
\centerline{{\bf{Abstract}}}
\vs{5}

We construct the string states $|O_{p}^J>_J$,  
$|O_{q}^{J_1}>_{{J_1}{J_2}}$  and  $|O_{0}^{J_{1}J_{2}}>_{{J_1}{J_2}}$  in
the Hilbert space of the quantum mechanical orbifold model
so as to calculate the three point functions and
the matrix elements of the light-cone Hamiltonian
from the interacting string bit model. With these string states
we show that the three point functions and
the matrix elements of the Hamiltonian
derived from the interacting string bit model 
up to $g^{2}_2$ order
precisely match with those computed
from the perturbative SYM theory in BMN limit.

\end{titlepage}
\newpage

\renewcommand{\thefootnote}{\arabic{footnote}}
\setcounter{footnote}{0}

\section{Introduction}

Recently, Berenstein, Maldacena, and Nastase (BMN) \cite{BMN} argued that the
IIB superstring theory on pp-wave background with RR-flux is dual
to a sector of ${\cal N}=4$ $SU(N)$ Yang-Mills thoery containing
operators with large $R$-charge $J$. The arguement was based on
the exact solvability of the Green-Schwarz strings on pp-wave
background obtained from $AdS_5\times S^5$ in the Penrose limit \cite{mets}.
In BMN limit, the effective 't Hooft coupling is $\la' = g_{YM}^2 N/{J^2}$.
Under this limit, the duality allows one to compute the free
string spectrum from the perturbative super Yang-Mills theory.
Moreover, it was shown in \cite{seme} and \cite{freed} that in BMN limit some
non-planar diagrams of arbitrary genus survive,
so besides the effective 't Hooft coupling is $\la'$, the string
interactions in pp-wave also involve genus parameter $g_2 = {J^2}/N$.
In \cite{freed}, it was proposed that the interaction amplitude for
a single string to split into two strings (or two strings joining
into one string) is related to the three point function of
the corresponding operators in the dual CFT, and $g_2\sqrt{\la'}$
was identified with the effective coupling between a wide class
of excited string states on the pp-wave backgound. By this proposal, 
the authors in \cite{freed} computed the second order correction to 
the anomalous dimension of the BMN operator from free planar
three point functions and found exact agreement with the 
computation of the torus contributions to the two
point function. Other related discussions on the string interaction
on pp-wave backgound can be found in  \cite{bn}--\cite{ap1}.

More recently, Verlinde proposed a string bit model \cite{thorn}
for interacting
strings on the pp-wave background in terms of the supersymmetric
quantum mechanics with a symmetric product target-space \cite{ver}, 
\cite{ver1}.
In this interacting string bit model, the 't Hooft coupling is
$\la^{2} = g_{YM}^2 N$, but the effective string coupling is
identified with the genus parameter $g_2 = {J^2}/N$.
In \cite{ver}, Verlinde presented some evidence that this effective
interacting string bit model would reproduce the complete
perturbation expansion of the ${\cal N}=4$ SYM theory in the BMN
limit.  
In \cite{ver}, 
Verlinde made an assumption that the three-point function in the string
bit model should be identified with that in the free SYM theory,
so it would be interesting to see how we can use the
operetor $\Sigma$ to
verify this assumption, and  whether we can reproduce the results derived
from the perturbation expansion of the  ${\cal N}=4$
SYM theory in the BMN limit, which would give a consistency
check for the interacting string bit model on pp-wave
background.

Motivated by the above, in the present paper we 
construct the string states $|O_{p}^J>_J$,
$|O_{q}^{J_1}>_{{J_1}{J_2}}$  and  $|O_{0}^{J_{1}J_{2}}>_{{J_1}{J_2}}$ 
in
the Hilbert space of the quantum mechanical orbifold model
so as to calculate the three point functions and 
the matrix elements of the light-cone Hamiltonian
from the interacting string bit model. The Hilbert space of the quantum
mechanical orbifold model is decomposed into the direct sum of
the Hilbert spaces of the twisted sectors \cite{vafa}, and
each twisted sector describes the states of several
strings, so the construction of the string states $|O_{p}^J>_J$,
$|O_{q}^{J_1}>_{{J_1}{J_2}}$ and $|O_{0}^{J_{1}J_{2}}>_{{J_1}{J_2}}$ 
can be realized by the fact that the vacuum state of
a twisted sector corresponds to a ground state twist
operator. We show that the three point functions and 
the matrix elements of the light-cone Hamiltonian derived from the
interacting string bit model up to $g^{2}_2$ order
precisely match with those computed
from the perturbative SYM theory in BMN limit.  In our calculation, 
instead of assuming that the three-point function in the string bit model
is the same as that in the free SYM theory, we derive this by
exploiting the operator  $\Sigma$, that is, we carry
out our calculation from the first principles of the 
the quantum mechanical orbifold model.

The paper is organized as follows. In the next section, we review
some basic results for the interacting
string bit model. In Section 3, 
we develope some approach to consistently
construct the string states $|O_{p}^J>_J$,
$|O_{q}^{J_1}>_{{J_1}{J_2}}$  
and $|O_{0}^{J_{1}J_{2}}>_{{J_1}{J_2}}$ in
the Hilbert space of the quantum mechanical orbifold model. 
In Section 4, the three point functions and 
the matrix elements between single and double
string states at the order $g_2$ and the matrix element
between two single string states at the order $g_{2}^2$
are calculated from the interacting string bit model.
In Section 5, we present our summary and discussion.

\section{The interacting string bit model}

Let us recapitulate some basic results for the interacting
string bit model \cite{ver}. One can introduce $J$ copies of
supersymmetric phase space coordinates 
$\{ p_n^i, x_n^i, \theta_n^a, \thetat_n^a\}$, with $n=1, \ldots, J$,
satisfying canonical commutation relations
\bea
[\, p^i_n , x^j_m\, ]  =  i\delta^{ij} \delta_{mn} \, , \qquad \ \ 
\{ \theta^a_n, \theta^b_m \}  =  \frac{1}{2}\delta^{ab} \delta_{mn} \, , 
\qquad \ \ 
\{ \thetat^a_n, \thetat^b_m \}  = \frac{1}{2}\delta^{ab} \delta_{mn} 
\label{cr} ~.
\eea
These $J$ copies can be regarded as obtained by the quantization
of the $J$-th symmetric product 
${\rm Sym}_J {\cal M}$ of the plane wave target space ${\cal M}$.
The Hilbert space of this quantum mechanical orbifold can be decomposed
into the direct sum of ``twisted sectors''
\bea
{\cal H} = \bigoplus_{[\gamma]} {\cal H}_{[\gamma]}
\eea
labeled by conjugacy classes $[\gamma]$ of the symmetric group
$S_J$ described by
\beq
[\gamma] = (1)^{J_1}(2)^{J_2}\ldots (s)^{J_s}
\eeq
where ${J_n}$ is the multiplicity of the cyclic permutation $(n)$
of $n$ elements. In each twisted sector, one should
keep only the states invariant under the centralizer
subgroup $C_g$ of $g$
\beq
C_g = \prod_{n=1}^{s} S_{J_n}\times Z_{n}^{J_n}
\eeq
where each factor $S_{J_n}$ permutes the ${J_n}$ cycles
$(n)$, while each  $Z_{n}$ acts within one particular
cycle $(n)$.

The Hilbert space ${\cal H}_{[\gamma]}$ of each twist sector
can be decomposed into the grade ${J_n}$-fold symmetric tensor
products of the Hilbert speces ${\cal H}_{n}$ which
correspond to the cycles of length $n$
\beq
{\cal H}_{\{J_n\}} =\bigotimes_{n=1}^{s} 
S_{J_n}{\cal H}_{n}=\bigotimes_{n=1}^{s}
\Big(\underbrace{
{\cal H}_{n}\bigotimes{\cal H}_{n}\ldots \bigotimes{\cal H}_{n}}_{{J_n\, 
times}}\Big)^{S_{J_n}}
\eeq
where the space  ${\cal H}_{n}$ is  $Z_{n}$ invariant subspace
of the  Hilbert space of the quantum mechanical orbifold
model of $16n$ bosonic fields 
$p_n^i$ and  $x_n^i$, and $16n$ fermionic
fields $\theta_n^a$ and $\thetat_n^a$.
The resulting  Hilbert space of the quantum mechanical orbifold
model is a sum over multi-string 
Hilbert spaces \cite{motl}--\cite{dvv}.

Consider the operator $\Sigma_{mn}$ that implement a simple
transposition of two string bits via
\beq
\Sigma_{nm} X_m = X_n 
\, \Sigma_{nm}\, , \qquad \qquad
\Sigma_{nm}   X_k = X_k \, \Sigma_{nm} \, \qquad 
\mbox{\footnotesize $k \neq m,n$}
\label{sigma}
\eeq
with $X_n = \{ p_n^i, x_n^i, \theta_n^a, \thetat_n^a\}$. By
acting with $\Sigma_{mn}$ on a given multi-string
sector, we get a different  multi-string sector via \cite{vafa}
\beq
\Sigma_{mn}: {\cal H}_\gamma \rightarrow{\cal H}_{\tilde\gamma}
\qquad \quad  \mbox{\footnotesize{with ${\tilde\gamma} = \gamma\cdot (mn)$}}
\label{hilb}  ~.
\eeq
When two sites $m$ and $n$ in the sector $\gamma$
correspond to one single string with length $J$ or two separate ones
with lengths $J_1$ and $(J-J_1)$, 
the new sector  $\tilde{\gamma}$ corresponds to either splitting the single 
string in two pieces of length $(m - n)$ and $(J -m + n)$, or 
joining the two strings to one of length $J$.

The light-cone supersymmetry generators and Hamiltonian of 
the free string theory are \cite{ver},\cite{ver1}
\beq
\label{expa}
Q_0 = Q_\da^{(0)} + \lambda\, Q_\da^{(1)} \ , \qquad 
\ \  
{\tilde{Q}}_0 = {\tilde{Q}}_\da^{(0)} - 
\lambda\, {\tilde{Q}}_\da^{(1)} \ , \qquad 
\ \  
H_0 = H^{(0)} + \lambda \, H_{\gamma}^{(1)} + \lambda^2  H_{\gamma}^{(2)}
\eeq
with
\bea
Q^{(0)}_\da = \sum_n (p_{n}^{i} {\gamma}_i\, \theta_n -  x^{i}_n({\gamma}_i
\Pi) \thetat_n)\, , \qquad 
Q_\da^{(1)} = \sum_n (x^i_{\gamma(n)} - x^i_{n})
{\gamma}_i\, \theta_n \nn
H^{(0)} = \sum_n (  \frac{1}{2 } (p_{i,n}^2 +
 x_{i,n}^2) + 2{\rm i}\, \thetat_n\Pi
\theta_n) \, , \qquad 
H^{(1)}_{\gamma} = - \sum_n 
{\rm i}(\theta_n  \theta_{\gamma(n)} 
-\thetat_n\thetat_{\gamma(n)} )
\label{Q}
\eea
\beq
H^{(2)}_{\gamma}
=  \sum_n \frac{1}{2 } (x^{i}_{\gamma(n)} - x^i_n)^2
\label{H}
\eeq
and\footnote{From \cite{BMN},\cite{freed} and \cite{gross},
we know that for free string, the $\lambda^2$ is identified
as $\frac{gN}{2\pi}$ with $g^{2}_{YM} = 4\pi\,g$, so the parameter
$\lambda^2$ should be $\frac{g^{2}_{YM}\,N}{8\pi^2}$, which
is different from that in \cite{ver} with extra factor 
$\frac{1}{8\pi^2}$.}
\beq
\lambda^2 = \frac{g^{2}_{YM}\,N}{8\pi^2} ~.
\eeq

The inner product that realizes the
combinatorics of the free gauge theory amplitudes in the string
bit language takes the form \cite{ver1}
\beq 
\langle \, \psi_1 | \,
\psi_2 \rangle_{g_2} = \, \langle \, \psi_1 | \, S \, | \, \psi_2
\rangle_{0} 
\eeq 
where $S$ is defined as $S = e^{g_2 \Sigma}$  with
\beq
\Sigma \equiv \, {1\over J^2} \sum_{m<n}\Sigma_{mn}\,~.
\label{sum}
\eeq
The free supersymmetry
generators can be split into two terms \cite{ver1}
\beq 
Q_0\, =
Q_{0}^{>} + Q_{0}^{<}
\label{split}
\eeq
and the general supersymmetry generators to all orders in ${g_2}$
are assumed to be 
\beq 
Q_0\, =
Q_{0}^{>} + S^{-1}Q_{0}^{<}S
\label{sg}
\eeq
where the $>$ superscript indicates the terms that contain
fermionic annihilation operators only, while $<$ denotes
terms with only fermionic creation operators.
The interacting light-cone Hamiltonian can be
extracted from the supersymmetry algebra
\beq 
\delta^{IJ}\, \{Q_{I}^{\dot{a}}, Q_{J}^{\dot{b}}\}
= \delta^{{\dot{a}}{\dot{b}}}  H \, +\, J^{\dot{a}\dot{b}}\,
\eeq 
where $J^{\dot{a}\dot{b}}$ is a suitable contraction of gamma matrices
with the $SO(4) \times SO(4)$ Lorentz generators $J^{ij}$ \cite{mets}.

The matrix elements of $H$ can be defined by \cite{ver1}
\beq 
\langle \psi_2 | (\delta^{\da\db} H + J^{\da\db})| \psi_1
\rangle_{g_2} = \delta^{{IJ}}\! \langle \psi_2 | \, S \, \{
Q_I^\da, Q_J^\db\}|\psi_1\rangle_0 
\eeq
from which the first and second order interaction terms can be
read off
\beq
H_1\,=g_2\,(U_1\,+U_2\,)\, , \qquad \qquad
H_2\,=g_2^2\,(V_1\,+V_2\,+V_3\,)
\label{h12}
\eeq
with
\beq
U_1\,= H_0\, \Sigma + \Sigma H_0\, , \qquad
U_2\,= -Q_0^>\Sigma\,Q_0^<\,
\label{u12}
\eeq
and
\beq
V_1\,= \frac{1}{2}(H_0\, \Sigma^2\, + \Sigma^2\, H_0\,) , \qquad 
V_2\,= -\frac{1}{2}Q_0^>\Sigma^2\,Q_0^<\, , \qquad 
V_3\,=[Q_0^>\, ,\Sigma][\Sigma\, ,Q_0^<\,]
\label{v123}
\eeq
Here we should point out that $H_1$ and $H_2$ are derived
from the bosonic matrix elements of $H$ \cite{ver1}.

\section{Construction of the string states in the interacting string bit model}

To construct the twisted vacuum states, let us first introduce
the ground state twist operator $\Sigma_{(n)}$ that translates 
the string bit $X_{I}$
within the individual string $(n)$ by one unit
\beq 
\Sigma_{(n)}   X_I = X_{I+1} \,\Sigma_{(n)} \, \qquad 
\mbox{\footnotesize $I=1,2,\ldots, n$} ~.
\eeq
The untwisted vacuum state $|0>$ is defined by
\beq 
a_{m}^{i} |0>=0\, , \ \ \qquad \
\beta_{m}^{a}|0>=0
\eeq
with 
\bea
x_{m}^{i}\! = \frac{1}{\sqrt{2}}(a_{m}^{i}\, + a_{m}^{i\,+})
\qquad \qquad \; {p_{m}^{i}}\, =\, \frac{{\rm i}}{\sqrt{2}}(a_{m}^{i}\,
-
a_{m}^{i\,+}) \nonumber \\[2.5mm]
\theta_n \! = \! \frac{1}{2}\left( \beta_n+\beta_n{}^+\right)
\qquad \qquad \Pi\theta_n=\frac{i}{2}\left(
\tilde{\beta}_n-\tilde{\beta}_n{}^+\right)
\\[2.5mm]
\tilde\theta_n\! = \! \frac{1}{ 2}\left(\tilde \beta_n{}
+\tilde\beta^{+}_n\right) \qquad \qquad
\tilde\theta_n\Pi=\frac{i}{ 2}\left(
\beta_n{}-\beta^{+}_n\right)\, . \nonumber 
\eea
and the untwisted vacuum state $|0>$ is normalized as
\beq 
<0|0>=1 ~.
\eeq
Then the twisted vacuum state $|n>$ is defined as
\beq 
|n> = \Sigma_{(n)}\, |0>
\eeq
with
\beq 
a_{m}^{i}\,|n>=0\, , \ \ \qquad \
\beta_{m}^{a}\,|n>=0\, , \ \ \qquad \
<n|n>=1
\eeq
which can be easily seen from the definition of $\Sigma_{(n)}$.

The arbitrary group element $\gamma\in\,S_J$ has the decomposition
\beq 
(n_1)(n_2)\ldots\,(n_l)
\eeq
where each cycle of length $n_\alpha$ has a definite set of indices
ordered up to a cyclic permutation and generates the action of
the subgroup $Z_{n_\alpha}$. Due to this decomposition, the 
operator  $\Sigma_{(\gamma)}$, $V_{\gamma}$ can be expressed
\beq
\Sigma_{(\gamma)}=\prod_{\alpha\,=1}^{l}\Sigma_{(n_\alpha)}\, , \ \ \qquad \
V_{\gamma}=\prod_{\alpha\,=1}^{l}V_{(n_\alpha)}
\eeq
where $V_{\gamma}$ is the operator defined in the twisted
sector ${\cal H}_{\gamma}$ .

To define an invariant operator  $V_{[\gamma]}$, we first
introduce the operator  $V_{\gamma}$ corresponding to a
fixed element $\gamma\in\,S_J$. Under the group action,
 $V_{\gamma}$ transforms into  $V_{h^{-1}\gamma\,h}$ \cite{af}, so
the invariant operator  $V_{[\gamma]}$ is the sum over
all operators from a given conjugacy class
\beq
V_{[\gamma]}=\frac{1}{J!}\sum_{h\in\,S_J}V_{h^{-1}\gamma\,h} ~.
\eeq
Any correlation function of the operators invariant under
centralizer subgroup $C_g$ should be invariant with respect
to the global action of the symmetric group \cite{af}
\beq
<V_{g_1}V_{g_2}\ldots\,V_{g_l}>=<V_{h^{-1}g_{1}h}V_{h^{-1}g_{2}h}
\ldots\,V_{h^{-1}g_{l}h}>
\label{ga}
\eeq
and the correlation function
\beq
<V_{g_1}V_{g_2}V_{g_3}>
\eeq
does not vanish only if\footnote{We should also include 
$g_1\,g_2\,g_3\,=1$, however, the action of the operator $\Sigma_{mn}$
on the Hilbert space is defined 
$\Sigma_{mn}: {\cal H}_\gamma \rightarrow{\cal H}_{\tilde\gamma}$
with ${\tilde\gamma} = \gamma\cdot (mn)$, so only $g_3\,g_2\,g_1\,=1$
is selected \cite{ver}.}
\beq
g_3\,g_2\,g_1\,=1 ~.
\label{321}
\eeq

The one-string state operator $O_{p}^{J}$ considered in \cite{BMN}
\beq
O_{p}^{J}=\frac{1}{\sqrt{JN^{J+2}}}\sum_{l=1}^{J}e^{2\pi\,ipl/J}
Tr(\phi\,Z^{l}\psi\,Z^{J-l})
\eeq
should be identified in the interacting string bit model
as \cite{ver}
\beq
O_{p}^{J}=\frac{1}{J}\Big(\sum_{k=1}^{J}a_{k}^{+} e^{-2\pi\,ipk/J}\Big)
\Big(\sum_{l=1}^{J}b_{l}^{+} e^{2\pi\,ipl/J}\Big)
\label{pj}
\eeq
which describes the one-string state and is invariant under
the centralizer subgroup $Z_J$. 

To construct the invariant state $|O_{p}^{J}>$, we introduce
the operator $O_{p,\gamma_1}^{J}$
\beq
O_{p,\gamma_1}^{J}=\frac{1}{J}\Big(\sum_{k=1}^{J}a_{\gamma_{1}(k)}^{+} 
e^{-2\pi\,ipk/J}\Big)
\Big(\sum_{l=1}^{J}b_{\gamma_{1}(l)}^{+} e^{2\pi\,ipl/J}\Big)
\label{o1}
\eeq
where $\gamma_1$ indicates one-cycle, i.e., one-string
state. The operator $O_{p,\gamma_1}^{J}$ is invariant under
the transformation of the centralizer subgroup $Z_J$,
and normalized as
\beq
<0|{O_{p,\gamma_1}^{J+}}\,O_{p,\gamma_1}^{J}|0>=1 ~.
\eeq
Then the invariant single string state  $|O_{p}^{J}>$
can be defined as 
\beq
|O_{p}^{J}>=\frac{\alpha}{J!}\sum_{h\in\,S_J}O_{p,h^{-1}{\gamma_1}\,h}^{J}
\Sigma_{(h^{-1}{\gamma_1}\,h)}|0>
\label{pstate}
\eeq
where $\alpha$ is the normalization factor which can be determined
by
\beq
<O_{p}^{J}|O_{p}^{J}>=1 ~.
\label{n1}
\eeq
By the centralizer subgroup $Z_J$, the normalization of the
state $|O_{p}^{J}>$ determines
\beq
\alpha = \sqrt{\frac{J!}{J}} ~.
\eeq

The two-string state operator  $O_{q,\gamma_2}^{J_1}$
can be constructed as
\beq
O_{q,\gamma_2}^{J_1} = \frac{1}{J_1}\Big(\sum_{k=1}^{J_1}a_{\gamma_{2}(k)}^{+} 
e^{-2\pi\,iqk/{J_1}}\Big)
\Big(\sum_{l=1}^{J_1}b_{\gamma_{2}(l)}^{+} e^{2\pi\,ipl/{J_1}}\Big)
\eeq
where $\gamma_2$ is decomposed as $(J_1)(J_2)$ with $J_2\,=J-J_1$,
that is, two cycles, and the operator $O_{q,\gamma_2}^{J_1}$
is invariant under the transformation of the centralizer subgroup
$Z_{J_1}\bigotimes\,Z_{J_2}$. Then the invariant two-string
state $|O_{q}^{J_1}>$ can be described as
\beq
|O_{q}^{J_1}>=\frac{\beta}{J!}\sum_{h\in\,S_J}O_{q,h^{-1}{\gamma_2}\,h}^{J_1}
\Sigma_{(h^{-1}{\gamma_2}\,h)}|0> .
\label{o2}
\eeq
The normalization of the state $|O_{q}^{J_1}>$ gives
\beq
\beta = \sqrt{\frac{J!}{J_{1}\,(J-J_{1})}}
\eeq
where we have used the centralizer subgroup $Z_{J_1}\bigotimes\,Z_{J_2}$.
The two-string state (\ref{o2}) corresponds to the state in ${\cal N}=4$
SYM theory \cite{BMN} and \cite{freed}
\beq
\frac{1}{J_{1}\,N^{J_{1}+2}}\sum_{l=1}^{J_1}e^{2\pi\,iql/{J_{1}}}
Tr(\phi\,Z^{l}\psi\,Z^{J_{1}\,-l})|0> ~.
\eeq

The other type two-string state operator $O_{0,\gamma_2}^{J_{1}\,J_{2}}$
can be constructed by
\beq
O_{0,\gamma_2}^{J_{1}\,J_{2}} = \frac{1}{\sqrt{J_{1}\,(J-J_{1})}}
\Big(\sum_{k=1}^{J_1}a_{\gamma_{2}(k)}^{+}\Big)
\Big(\sum_{l=J_{1}\,+1}^{J}b_{\gamma_{2}(l)}^{+}\Big)
\label{o3}
\eeq
which is invariant under the transformation of the 
centralizer subgroup $Z_{J_1}\bigotimes\,Z_{J_2}$, and
normalized as
\beq
<0|O_{0,\gamma_2}^{J_{1}\,J_{2}+}O_{0,\gamma_2}^{J_{1}\,J_{2}}|0>=1 ~.
\eeq
The corresponding normalized and invariant two-string state is
\beq
|O_{0}^{J_{1}\,J_{2}}> = \frac{\beta}{J!}\sum_{h\in\,S_J}O_{0,h^{-1}
{\gamma_2}\,h}^{J_1\,J_2}
\Sigma_{(h^{-1}{\gamma_2}\,h)}|0>
\eeq
where the centralizer subgroup 
$Z_{J_1}\bigotimes\,Z_{J_2}$ is exploited. The two-string
state $|O_{0}^{J_{1}\,J_{2}}>$ corresponds to the 
following state in ${\cal N}=4$
SYM theory \cite{BMN} and \cite{freed}
\beq
\frac{1}{N^{J+2}}
Tr(\phi\,Z^{J_{1}})\,Tr(\psi\,Z^{J-J_{1}})|0> ~.
\eeq

Up to now, we have constructed the consistent string states 
$|O_{p}^J>_J$,  
$|O_{q}^{J_1}>_{{J_1}{J_2}}$  and  $|O_{0}^{J_{1}J_{2}}>_{{J_1}{J_2}}$ 
in the interacting string bit model. In the next section,
we will use them to calculate the three point functions and the
matrix elements of the light-cone Hamiltonian in $g_2$ and
$g_2^2$ order.

\section{Three point functions and 
matrix elements of the light-cone Hamiltonian in $g_2$ and
$g_2^2$ order}

Exploiting the above constructed string states, we first
consider the three point function between the single
string state $|O_{p}^{J}>$ and the two-string state $|O_{q}^{J_1}>$
induced by the action $\Sigma$
\beq
C_{pqx}=\,{}_{\strut J}\! \langle \,O_{p}^{J}|\Sigma\,|O_{q}^{J_1}
\,\rangle_{\strut{\! J_1 J_2}} 
\label{3i}
\eeq
with $x=J_{1}/J$.

Inserting (\ref{sum}), (\ref{o1}) and  (\ref{o2}) into  (\ref{3i}), we have
\bea
C_{pqx}&=&\frac{\alpha\beta}{(J!)^2}
\sum_{h,h^{\prime}\in\,S_J} \langle\,0|\Sigma_{(h^{-1}\gamma_{1}h)}
O_{p,h^{-1}\gamma_{1}h}^{J+}\Sigma\,\nn
&\cdot\,&
O_{q, {h^{\prime}}^{-1}\gamma_{2}
h^{\prime}}^{J_1}\Sigma_{ ({h^{\prime}}^{-1}\gamma_{2}
h^{\prime})}|0\rangle ~.
\label{pq1}
\eea
By exploiting (\ref{ga}), the 
three point function  (\ref{pq1}) can be recast into
\bea
C_{pqx}&=&\frac{1}
{J^2\sqrt{JJ_{1}(J-J_{1})}}
\sum_{h\in\,S_J}\sum_{m<n} \langle\,0|\Sigma_{(h^{-1}\gamma_{1}h)}
O_{p,h^{-1}\gamma_{1}h}^{J+}\nn
&\cdot\,&
\Sigma_{mn}\,
O_{q, \gamma_{2}}^{J_1}
\Sigma_{ (\gamma_{2})}|0\rangle ~.
\eea
Since the state $|O_{q}^{J_1}
\,\rangle_{\strut{\! J_1 J_2}}$ is two-string state,
to obtain the single string state $|O_{p}^{J}
\,\rangle_{\strut{\! J}}$, the variables $m$ and $n$
should take the following values
\beq
1\leq\, m\leq\,J_1,\, , \qquad 
J_1\,+1\leq\, n\leq\,J
\eeq
and the above three-point function can be rewritten as
\bea
C_{pqx} &=& \frac{g_{2}\lambda^{2}}
{J^2\sqrt{JJ_{1}(J-J_{1})}}
\sum_{h\in\,S_J}\sum_{m=1}^{J_1}\sum_{n=J_1\,+1}^{J} 
\langle\,0|\Sigma_{(h^{-1}\gamma_{1}h)}
O_{p,h^{-1}\gamma_{1}h}^{J+}\nn
&\cdot\,&
\Sigma_{mn}\,O_{q, \gamma_{2}}^{J_1}
\Sigma_{ (\gamma_{2})}|0\rangle ~.
\label{pq2}
\eea

Applying the centralizer subgroup $Z_{J_1}\bigotimes\,Z_{J_2}$,
Eq. (\ref{pq2}) can be reduced to 
\bea
C_{pqx} &=& \frac{J_{1}(J-J_{1})}
{J^2\sqrt{JJ_{1}(J-J_{1})}}
\sum_{h\in\,S_J}
\langle\,0|\Sigma_{(h^{-1}\gamma_{1}h)}
O_{p,h^{-1}\gamma_{1}h}^{J+}\nn
&\cdot\,&
\Sigma_{J_{1}J}\,O_{q, \gamma_{2}}^{J_1}
\Sigma_{ (\gamma_{2})}|0\rangle ~.
\eea
By the centralizer subgroup $Z_{J}$ and the condition (\ref{321}),
we arrive at
\bea
C_{pqx} &=& \frac{1}{J}
\sqrt{\frac{J_{1}(J-J_{1})}{J}}
\langle\,0|\Sigma_{(\gamma_{1})}
O_{p,\gamma_{1}}^{J+}\nn
&\cdot\,&
\Sigma_{J_{1}J}\,O_{q, \gamma_{2}}^{J_1}
\Sigma_{ (\gamma_{2})}|0\rangle
\label{pqf}
\eea
with
\beq
\c_{2}\c_{J_{1}\,J}\c_{1}=1 ~.
\label{unit}
\eeq

To simplify the following the calculation, we choose
\beq
\c_{1}=(1,2,3,\ldots\,J-1,J)
\label{c1}
\eeq
then (\ref{unit}) gives
\beq
\c_{2}^{-1}=\c_{J_{1}\,J}\cdot\,\c_{1}=\Big(\underbrace{
1 2 3\ldots\, J_{1}-2, J_{1}-1, J}_{{J_1\, 
times}}\Big)\Big(\underbrace{J_{1}, J_{1}+1,\ldots\, J-2, J-1
}_{{J-J_1\, 
times}}\Big)
\label{c2}
\eeq
and the operators $O_{p,\gamma_{1}}^{J+}$ and $O_{q, \gamma_{2}}^{J_1}$
are simply given by
\bea
O_{p,\gamma_{1}}^{J+}&=&\frac{1}{J}\Big(\sum_{m=1}^{J}a_{m}^{+} 
e^{2\pi\,ipm/J}\Big)
\Big(\sum_{l=1}^{J}b_{l}^{+} e^{-2\pi\,ipl/J}\Big)\nn
O_{q, \gamma_{2}}^{J_1}&=&\frac{1}{J_1}\Big(\sum_{m=1}^{J_1\,-1}a_{m}^{+} 
e^{-2\pi\,iqm/{J_1}} + a_{J}^{+} \Big)
\Big(\sum_{l=1}^{J_1\,-1}b_{l}^{+} e^{2\pi\,iql/{J_1}} + b_{J}^{+} \Big)
\label{opq}
\eea
where we have used the fact that the operators
$O_{p,\gamma_{1}}^{J+}$ and $O_{q, \gamma_{2}}^{J_1}$
are invariant under the transformation of the centralizer
subgroup $Z_J$, $Z_{J_1}\bigotimes\,Z_{J_2}$ respectively.

Plugging (\ref{opq}) into  (\ref{pqf}), we get
\beq
C_{pqx} =\frac{1}{J^{2}}\sqrt{\frac{J-{J_1}}{J{J_1}}}
\frac{\sin^{2}\pi\,px}
{\sin\pi(\frac{p}{J}-\frac{q}{J_1})}
\label{exact}
\eeq
where we have used the relation
$\langle\, 0|\Sigma_{\c_1}\Sigma_{J_{1}J}\Sigma_{{\c_1}^{-1}\c_{J_{1}J}}
|0\rangle\,=1$ which can be derived by (\ref{hilb}) and (\ref{unit}).

In the BMN limit, $\frac{p}{J}$, $\frac{q}{J_1}$ are very
small \cite{BMN}, so (\ref{exact}) can be written as
\beq
C_{pqx} =
\sqrt{\frac{1-x}{Jx}}\frac{\sin^{2}\pi\,px }
{\pi^2\, {(p-\frac{q}{x})}^2\,}
\label{apr}
\eeq
which agrees with the three point function calculated from
the perturbative SYM theory \cite{freed}.
Here we should stress that the three-point function calculated
from the interacting string bit model is (\ref{exact}).
Only in the case of the small $\frac{p}{J}$ and $\frac{q}{J_1}$,
the (\ref{exact}) can be rewritten as (\ref{apr}).

The other three-point function including
the operator $O_{0,\c_2}^{J_{1}J_{2}}$ can be calculated 
in the way like for $C_{pqx}$
\bea
C_{px} &=&{}_{\strut J}\! \langle \,O_{p}^{J}|\Sigma|O_{0}^{J_1\,J_2}
\,\rangle_{\strut{\! J_1 J_2}}\nn
&=&
\frac{\alpha\beta}{(J!)^2}
\sum_{h,h^{\prime}\in\,S_J} \langle\,0|\Sigma_{(h^{-1}\gamma_{1}h)}
O_{p,h^{-1}\gamma_{1}h}^{J+}\Sigma\nn
&\cdot\,&
O_{0, {h^{\prime}}^{-1}\gamma_{2}
h^{\prime}}^{J_1\,J_2}\Sigma_{ ({h^{\prime}}^{-1}\gamma_{2}
h^{\prime})}|0\rangle .
\eea
In the similar way, we have
\bea
C_{px}&=&g_{2}\lambda^{2}\frac{1}{J}
\sqrt{\frac{J_{1}(J-J_{1})}{J}}
\langle\,0|\Sigma_{(\gamma_{1})}
O_{p,\gamma_{1}}^{J+}\nn
&\cdot\,&
\Sigma_{J_{1}J}O_{0, \gamma_{2}}^{J_1\,J_2}
\Sigma_{ (\gamma_{2})}|0\rangle ~.
\label{pp}
\eea
Inserting  (\ref{sum}), (\ref{opq}), (\ref{o3}), (\ref{c1}), (\ref{c2}) 
into (\ref{pp}), we obtain that
\beq
C_{px}={}_{\strut J}\! \langle \,O_{p}^{J}|\Sigma|O_{0}^{J_1\,J_2}
\,\rangle_{\strut{\! J_1 J_2}}=\frac{\sin^{2}\pi\,px}{\pi^2\,\sqrt{J}\,p^2}
\label{pp1}
\eeq
which is exactly the same as that derived from the
perturbative SYM theory \cite{freed}.

The matrix element of $\Sigma^2$ between the single
string states represents the one-loop contribution
due to successive splitting and joining, which can be
obtained by factorization
\beq
\,{}_{\strut J}\! \langle \,O_{p}^{J}|\Sigma^2\,|O_{q}^{J}
\,\rangle_{\strut{\! J}} = \sum_{k,x}\, C_{pkx}C_{qkx}
+  \sum_{x}\, C_{px}C_{qx} = 2A_{pq}
\label{fac}
\eeq
where the explicit form of $A_{pq}$ is given in  \cite{freed}
and \cite{seme}.

Now let us consider the matrix element between single and double
string states at $g_2$ order, which corresponds to an operator
mixing term in the gauge theory. From 
(\ref{h12}), we have
\beq
\,{}_{\strut J}\! \langle \,O_{p}^{J}|H_1\,|O_{q}^{J_1}
\,\rangle_{\strut{\! J_1 J_2}} = g_2\,
\,{}_{\strut J}\! \langle \,O_{p}^{J}|U_1\,|O_{q}^{J_1}
\,\rangle_{\strut{\! J_1 J_2}} + g_2\,
\,{}_{\strut J}\! \langle \,O_{p}^{J}|U_2\,|O_{q}^{J_1}
\,\rangle_{\strut{\! J_1 J_2}}
\label{gi}
\eeq
where $U_1$ and $U_2$ are defined in (\ref{u12}).

The string states $|O_{p}^J>_J$,  
$|O_{q}^{J_1}>_{{J_1}{J_2}}$  and  $|O_{0}^{J_{1}J_{2}}>_{{J_1}{J_2}}$ 
are eigenstates of the free Hamiltonian  $H_0$, the first term
in (\ref{gi}) can be easily obtained by exploiting (\ref{apr})
\beq
 g_2\,
\,{}_{\strut J}\! \langle \,O_{p}^{J}|U_1\,|O_{q}^{J_1}
\,\rangle_{\strut{\! J_1 J_2}} = g_2\,\la'\,(p^2\,+\frac{q^2}{x^2})
C_{pqx}
\label{u1}
\eeq
where $C_{pqx}$ is the three point function defined in (\ref{apr}).

To calculate the second term in (\ref{gi})
\beq
\,{}_{\strut J}\! \langle \,O_{p}^{J}|U_2\,|O_{q}^{J_1}
\,\rangle_{\strut{\! J_1 J_2}}=
-g_2\,\,{}_{\strut J}\! \langle \,O_{p}^{J}|Q^{>}_{\gamma_{1}}\,\Sigma\,
Q^{<}_{\gamma_{2}}|O_{q}^{J_1}
\,\rangle_{\strut{\! J_1 J_2}}
\label{u2}
\eeq
we give the explicit
form for $Q^{>}_{\gamma_{1}}$ and $Q^{<}_{\gamma_{2}}$
\bea 
Q^{<}_{\gamma_1}&=& \lambda
\sum_{m=0}^{J-1}\bigg[(a_{m+1}^{i \dag}\! +a_{m+1}^i)-(a_m^{i\
\dag} \! +
a_m^i)\bigg]\gamma^i\beta_m^\dag\, \nn
Q^{>}_{\gamma_2}& = &\lambda \sum_{m=0}^{J-1}\beta_m\gamma^i
\Bigl[(a_{m+1}^{i\ \dag} \! +\! a_{m+1}^i)\! -\!
(a_m^i{}^\dag+a_m^i)\Bigr]\nn
&+&
\, (\beta_{{J_1\! -1}}\!\! -\!
\beta_{{J-1}})\gamma^i\bigg[(a_J^{i\ \dag}\! +\! a_J^i)
-(a_{J_1}^{i\ \dag}+ a_{J_1}^i)\bigg]\, 
\label{qq}
\eea
where we have identified
the last site {\small $m=J$} with the 0-th site {\small $m=0$}.
Then the second term can be recast into
\beq
-\,{}_{\strut J}\! \langle \,O_{p}^{J}|Q^{>}_{\gamma_{1}}\,\Sigma\,
Q^{<}_{\gamma_{2}}|O_{q}^{J_1}
\,\rangle_{\strut{\! J_1 J_2}}=X_1\, + X_2\,
\label{x12}
\eeq
with 
\bea 
X_1&=& -g_2\,\lambda^2\,
\,{}_{\strut J}\! \langle \,O_{p}^{J}|
\bigg(\sum_{m=1}^{J}(a_{m+1}^{i \dag}\! -a_{m}^{i \dag})\gamma^i\beta_m
\bigg)\Sigma\bigg( \sum_{n=1}^{J}\beta_n^\dag\,\gamma^i\,
(a_{n+1}^{i} -a_{n}^{i})\nn
&+&
(\beta_{J_1}^\dag\,-
\beta_{{J-1}}^\dag\,)\gamma^i\,(a_J^{i}-a_{J_1}^{i})
 \bigg)
|O_{q}^{J_1}
\,\rangle_{\strut{\! J_1 J_2}}\nn
X_2&=& -g_2\,\lambda^2\,
\,{}_{\strut J}\! \langle \,O_{p}^{J}|
\bigg(\sum_{m=1}^{J}(a_{m+1}^{i}\! -a_{m}^{i})\gamma^i\beta_m
\bigg)\Sigma\bigg(\sum_{n=1}^{J}\beta_n^\dag\,\gamma^i\,
(a_{n+1}^{i\dag} -a_{n}^{i\dag})\nn
&+&
(\beta_{J_1}^\dag\,-
\beta_{{J-1}}^\dag\,)\gamma^i\,(a_J^{i\dag}-a_{J_1}^{i\dag})
 \bigg)
|O_{q}^{J_1}
\,\rangle_{\strut{\! J_1 J_2}}~.
\label{x12d}
\eea
After taking the inner product and keeping track of the
action of the centralizers, we have\footnote{When we calculate
$X_2$ some contribution from normal ordering arises,
which corresponds to a vacuum fluctuation and can be
cancelled by the hopping terms in the Hamiltonian \cite{ver1}.}
\bea
X_1&=&-\frac{1}{2}g_2\,\la'\,\frac{pq}{x}\,C_{pqx}\nn
X_2&=&-\frac{1}{2}g_2\,\la'\,\frac{pq}{x}\,C_{pqx}
\label{x}
\eea
where (\ref{apr}) has been used in the calculation.
 
Inserting  (\ref{u1}), (\ref{x12}) and (\ref{x}) 
into (\ref{gi}), we have
\beq
\,{}_{\strut J}\! \langle \,O_{p}^{J}|H_1\,|O_{q}^{J_1}
\,\rangle_{\strut{\! J_1 J_2}} = g_2\,\la'\,\bigg(p^2\,+\frac{q^2}{x^2}
- \frac{pq}{x}\bigg)C_{pqx}~.
\label{r1}
\eeq
Similary, we have 
\beq
\,{}_{\strut J}\! \langle \,O_{p}^{J}|H_1\,|O_{0}^{J_{1}J_{2}}
\,\rangle_{\strut{\! J_1 J_2}} = g_2\,\la'\,p^2\,C_{px}~.
\label{r2}
\eeq
In gauge theory, the mixing between single and double trace
operators is  \cite{seme1}, \cite{freed1}
\beq
\langle \, O_{p}^{J}(0) :{\bar{O}}_{q}^{J_{1}}{\bar{O}}^{J_{1}}:(x)\rangle
= g_2\,C_{pqx}\bigg(1- \la'\,\ln(x\Lambda)^2\,(p^2\,+\frac{q^2}{x^2}
- \frac{pq}{x})\bigg)
\eeq
which precisely matches (\ref{r1}).

Finally, we calculate the matrix element between two single string
states at $g_2^2$ order
\beq
\,{}_{\strut J}\! \langle \,O_{p}^{J}|H_2|O_{q}^{J}
\,\rangle_{\strut{\! J}}=\,{}_{\strut J}\! \langle \,O_{p}^{J}|V_1|O_{q}^{J}
\,\rangle_{\strut{\! J}}+
\,{}_{\strut J}\! \langle \,O_{p}^{J}|V_2|O_{q}^{J}
\,\rangle_{\strut{\! J}}+
\,{}_{\strut J}\! \langle \,O_{p}^{J}|V_3|O_{q}^{J}
\,\rangle_{\strut{\! J}}
\label{g2i}
\eeq
where $V_1$, $V_2$ and $V_3$ are defined in (\ref{v123}).

From (\ref{u12}) and (\ref{v123}),
we see $V_1$ and $V_2$ have similar structure to $U_1$ and $U_2$
with replacing $\Sigma$ by $\Sigma^2$. By exploiting the
factorization (\ref{fac}), we find
\beq
{}_{\strut J}\! \langle \,O_{p}^{J}|V_1|O_{q}^{J}
\,\rangle_{\strut{\! J}}=g_2^2\,\la'\,
(p^2\,+q^2)A_{pq}, \qquad \ \ 
\,{}_{\strut J}\! \langle \,O_{p}^{J}|V_2|O_{q}^{J}
\,\rangle_{\strut{\! J}}=-g_2^2\,\la'\,pqA_{pq},
\label{v12}
\eeq
where $A_{pq}$ is the matrix element of the interaction
term $\Sigma^2$ defined in  (\ref{fac}). 

The matrix
element $\,{}_{\strut J}\! \langle \,O_{p}^{J}|V_3|O_{q}^{J}
\,\rangle_{\strut{\! J}}$ can be expressed by the factorization
\bea
\,{}_{\strut J}\! \langle \,O_{p}^{J}|V_3|O_{q}^{J}
\,\rangle_{\strut{\! J}}&=&g_2^2\,\sum_{r,J_{1}}
{}_{\strut J}\! \langle \,O_{p}^{J}|\Sigma|O_{q}^{J_1}
\,\rangle_{\strut{\! J_1 J_2}}\,\,{}_{\strut{\! J_1 J_2}}
\! \langle \,O_{r}^{J_1}|[Q_0^{>},\Sigma]Q_0^{<}|O_{q}^{J}
\,\rangle_{\strut{\! J}}\nn
&-&
g_2^2\,\sum_{r,J_{1}}
{}_{\strut J}\! \langle \,O_{p}^{J}|[Q_0^{>},\Sigma]Q_0^{<}
|O_{q}^{J_1}
\,\rangle_{\strut{\! J_1 J_2}}\,\,{}_{\strut{\! J_1 J_2}}
\! \langle \,O_{r}^{J_1}|\Sigma\,|O_{q}^{J}
\,\rangle_{\strut{\! J}}\nn
&+&
g_2^2\,\sum_{J_{1}}
{}_{\strut J}\! \langle \,O_{p}^{J}|\Sigma|O_{0}^{J_1\, J_2}
\,\rangle_{\strut{\! J_1 J_2}}\,\,{}_{\strut{\! J_1 J_2}}
\! \langle \,O_{0}^{J_1\, J_2}|[Q_0^{>},\Sigma]Q_0^{<}|O_{q}^{J}
\,\rangle_{\strut{\! J}}\nn
&-&
g_2^2\,\sum_{J_{1}}
{}_{\strut J}\! \langle \,O_{p}^{J}|[Q_0^{>},\Sigma]Q_0^{<}
|O_{0}^{J_1\, J_2}
\,\rangle_{\strut{\! J_1 J_2}}\,\,{}_{\strut{\! J_1 J_2}}
\! \langle \,O_{0}^{J_1\, J_2}|\Sigma\,|O_{q}^{J}
\,\rangle_{\strut{\! J}}
\label{v3e}
\eea
where in the large $J$,
we have used the relation $[\Sigma,[\Sigma, Q_0^{>}]]=0$.

By exploiting (\ref{u1}), (\ref{x12}) and (\ref{x}),
(\ref{v3e}) can be written as
\beq
\,{}_{\strut J}\! \langle \,O_{p}^{J}|V_3|O_{q}^{J}
\,\rangle_{\strut{\! J}}=g_2^2\,\la'\,\sum_{r,x}\bigg(
p(\frac{r}{x}-p)-\frac{r}{x}(r-\frac{r}{x})\bigg)C_{prx}C_{qrx}-
g_2^2\,\la'\,\sum_{x}\,p^2\,C_{px}C_{qx}=\frac{g_2^2\,\la'\,}{4\pi^2}B_{pq}
\label{v3}
\eeq
where the explicit form of $B_{pq}$ can be found in \cite{seme1}
and \cite{freed1}. 
Inserting (\ref{v12}) and (\ref{v3}) into (\ref{g2i}), we arrive
at
\beq
\,{}_{\strut J}\! \langle \,O_{p}^{J}|H_2|O_{q}^{J}
\,\rangle_{\strut{\! J}}=g_2^2\,\la'\,\bigg((p^2\,+q^2\,-pq)A_{pq}
+\frac{1}{4\pi^2}B_{pq}\bigg)
\label{r3}
\eeq
which agrees with the result obtained from perturbative
Yang-Mills calculation \cite{seme1},\cite{freed1}
\beq
\,{}_{\strut J}\! \langle \,O_{p}^{J}(0){\bar{O}}_{q}^{J}(x)
\,\rangle_{\strut{\! J}}=(\delta_{pq}\,+ g_2^2\,A_{pq})
\bigg[1-(p^2\,+q^2\,-pq)\la'\,\ln(x\Lambda)^2\, \bigg]
-\frac{g_2^2\,\la'}{4\pi^2}B_{pq}\ln(x\Lambda)^2\,~.
\eeq

From (\ref{apr}), (\ref{pp1}), (\ref{r1}), (\ref{r2}) and
(\ref{r3}),
we find that the three-point 
functions ${}_{\strut J}\! \langle \,O_{p}^{J}|\Sigma\,|O_{0}^{J_1\,J_2}
\,\rangle_{\strut{\! J_1 J_2}}$ and ${}_{\strut J}
\! \langle \,O_{p}^{J}|\Sigma\,|O_{0}^{J_1\,J_2}
\,\rangle_{\strut{\! J_1 J_2}}$, the matrix elements 
$\,{}_{\strut J}\! \langle \,O_{p}^{J}|H_1\,|O_{q}^{J_1}
\,\rangle_{\strut{\! J_1 J_2}}$,
$\,{}_{\strut J}\! \langle \,O_{p}^{J}|H_1\,|O_{0}^{J_{1}J_{2}}
\,\rangle_{\strut{\! J_1 J_2}}$, 
$\,{}_{\strut J}\! \langle \,O_{p}^{J}|H_2|O_{q}^{J}
\,\rangle_{\strut{\! J}}$
obtained from the
interacting string bit model precisely match to those
derived from the ${\cal\,N}=4$ SYM thoery.
Instead of identifying the three-point function in the string
bit model with those in the free SYM  theory \cite{ver},
in the above we have exploited the operator 
$\Sigma$ to carry out our calculation.

\section{Summary}

So far, we have developed new approach to consistently
construct the string states $|O_{p}^J>_J$,  $|O_{q}^{J_1}>_{{J_1}{J_2}}$
and  
$|O_{0}^{{J_1}{J_2}}>_{{J_1}{J_2}}$  in
the Hilbert space of the quantum mechanical orbifold model
in order to  calculate the
three-point functions, and the matrix elements of the
light-cone Hamiltonian from
the interacting string bit model.
Since the Hilbert space of the quantum
mechanical orbifold model can be decomposed into the direct sum of
the Hilbert spaces of the twisted sectors, and
each twisted sector describes the states of several
strings, the  construction of  the string states $|O_{p}^J>_J$,  
$|O_{q}^{J_1}>_{{J_1}{J_2}}$ and  
$|O_{0}^{{J_1}{J_2}}>_{{J_1}{J_2}}$
has  been realized by the fact that the vacuum state of
a twisted sector can be described by the ground state twist
operator. We have shown that for 
the three-point functions, and the matrix elements of the
light-cone Hamiltonian up to $g^2_2$ order, the results obtained by
interacting string bit model precisely match with those computed
from the perturbative SYM theory in BMN limit. 
We should emphasize that in our calculation, 
instead of assuming that the three-point functions in the string bit model
are the same as those in the free SYM theory, we have derived those
from the first principles of the 
the quantum mechanical orbifold model.

In (\ref{cr}), we have introduced $J$
copies of supersymmetric phase space coordinates, and the operator
$O_{p}^{J}$ is identified in the interacting string
bit model as (\ref{pj}). In \cite{seme} and 
\cite{freed}, it was shown that the sum should start from $l=0$ 
instead of $l=1$ and this
small difference has drastic consequences. In particular the operator with
sum beginning at $l=1$ does not reduce to a chiral primary for $p\rightarrow\,
0$. Thus the operator $O_{p}^{J}$ should be defined 
\beq
O_{p}^{J}=\frac{1}{J}\Big(\sum_{k=0}^{J}a_{k}^{+} e^{-2\pi\,ipk/J}\Big)
\Big(\sum_{l=0}^{J}b_{l}^{+} e^{2\pi\,ipl/J}\Big)~.
\label{pj1}
\eeq
One may wonder if our results would be changed by
introducing $J+1$ copies of supersymmetric phase space coordinates, 
and the operator $O_{p}^{J}$ instead defined as
(\ref{pj1}),
but after some calculation, we find that is not the case.
In the above, the interacting string bit model has been
constructed by the harmonic osillators.
It would be interesting  to construct
the interacting string bit model in terms of 
Cuntz osillators  \cite{gg}
and see whether both approaches match with each other.
We hope to return to these issues in near future.

\section*{Acknowledgments}

I thank M. Walton for useful discussion.
This work was supported by NSERC.


\begin{thebibliography}{99}

\bibitem{BMN}
D.~Berenstein, J.~Maldacena, H.~Nastase, {\sl Strings in flat
space and pp waves from N = 4 super Yang Mills}, 
hep-th/0202021.

\bibitem{mets}
R.~R.~Metsaev, {\sl Type IIB Green-Schwarz superstring in plane wave
Ramond-Ramond  background}, Nucl.\ Phys.\ B {\bf 625}, 70 (2002),
hep-th/0112044; R.~R.~Metsaev,  A.~A.~Tseytlin,
{\sl Exactly solvable model of superstring in plane wave
Ramond-Ramond  background}, hep-th/0202109.

\bibitem{seme}
C. Kristjansen, J. Plefka, G. W. Semenoff, M. Staudacher, {\sl A
New double-scaling limit of N=4 super Yang-Mills theory and
pp-wave strings}, hep-th/0205033.

\bibitem{freed}
N.R. Constable, D.Z. Freedman, M. Headrick, S. Minwalla, L. Motl,
A. Postnikov, W. Skiba, {\sl PP-wave string interactions from
perturbative Yang-Mills theory}, hep-th/0205089.

\bibitem{bn}
D. Berenstein and H. Nastase, {\sl On lightcone string field theory
from super Yang-Mills and holography},  hep-th/0205048.

\bibitem{sv1}
M. Spradlin, A. Volovich, {\sl Superstring interactions in a
pp-wave background}, hep-th/0204146.

\bibitem{gopakumar}
R. Gopakumar, {\sl String Interactions in PP-waves},
hep-th/0205174.


\bibitem{kklp}
Y. Kiem, Y. Kim, S. Lee, J. Park, {\sl PP-wave/Yang-Mills
Correspondence: An explicit check}, hep-th/0205279.

\bibitem{minxin}
M. Huang, {\sl Three point functions of N=4 super Yang Mills from
light cone string field theory in pp-wave}, hep-th/0205311;
{\sl String interactions in PP-wave from ${\cal N}=4$ super Yang-Mills},
hep-th/0206248.

\bibitem{chu1}
C. Chu, V.V. Khoze, G. Travaglini, {\sl Three-point functions in N=4
super Yang-Mills theory and pp-wave}, hep-th/0206005.

\bibitem{lmp}
P. Lee, S. Moriyama, J. Park, {\sl Cubic Interactions in PP-wave
light cone string field theory} , hep-th/0206065.

\bibitem{sv2}
M. Spradlin and A. Volovich,  {\sl Superstring interactions in a
pp-wave background II}, hep-th/0206073.

\bibitem{chu2}
C. Chu, V.V. Khoze, G. Travaglini, {\sl PP-wave string interactions
from n-point correlators of BMN operators},  hep-th/0206167.

\bibitem{new}
I. R. Klebanov, M. Spradlin, A. Volovich, {\sl New Effects in Gauge
Theory from pp-wave Superstrings}, hep-th/0206221.

\bibitem{ug}
U. Gursoy, {\sl Vector operators in the BMN correspondence}, hep-th/0208041.

\bibitem{chu3}
C. Chu, V.V. Khoze, M. Petrini, R. Russo, A. Tanzini, 
{\sl A note on string interaction on pp-wave
background}, hep-th/0208148.

\bibitem{bei}
N. Beisert, C. Kristjansen, J. Plefka, G.W. Semenoff, M. Staudacher,
{\sl BMN correlators and operator mixing in N=4 Super Yang-Mills theory}, 
hep-th/0208178.

\bibitem{schw}
 J. H. Schwarz, {\sl Comments on superstring interactions in a plane-wave 
background}, hep-th/0208179.

\bibitem{ap1}
A. Pankiewicz, {\sl More comments on superstring interactions 
in the pp-wave background}, hep-th/0208209.

\bibitem{thorn}
R. Giles, C.~B.~Thorn,{\sl A Lattice Approach To String Theory},
Phys.\ Rev.\ D {\bf 16}, 366 (1977);
C.~B.~Thorn, {\sl
Supersymmetric quantum mechanics for
string-bits}, Phys.\ Rev.\ D {\bf 56}, 6619 (1997), hep-th/9707048;
C.~B.~Thorn,{\sl
A Fock Space Description Of The 1/N-C Expansion Of Quantum Chromodynamics},
Phys.\ Rev.\ D {\bf 20}, 1435 (1979).

\bibitem{ver}
H. Verlinde, {\sl Bits, matrices and 1/N}, hep-th/0206059.

\bibitem{ver1}
D. Vaman, H. Verlinde, {\sl Bit strings from N=4 gauge theory}, 
hep-th/0209215.

\bibitem{vafa}
C. Vafa, E. Witten, {\sl Strong coupling test of S-duality}, 
Nucl.Phys. {\bf B431} (1994) 3, hep-th/9408074.


\bibitem{motl}
L. Motl, {\sl Proposals on nonpertubative superstring interactions},
hep-th/9701025.

\bibitem{bank}
T. Banks, N. Seiberg, {\sl Strings from
matrices},  Nucl. Phys. {\bf B497} (1997) 41, hep-th/9702187.

\bibitem{dvv}
R. Dijkgraaf, E. Verlinde and H. Verlinde, {\sl Matrix string theory},
Nucl. Phys. B {\bf 500} (1997) 43, hep-th/9703030.

\bibitem{bon}
G. Bonelli, {\sl Matrix strings in pp-wave backgrounds from deformed 
super Yang-Mills theory}, hep-th/0205213.

\bibitem{gross}
D.J. Gross, A. Mikhailov, R. Roiban, {\sl  Operators with large R
charge in N=4 Yang-Mills theory}, hep-th/0205066.

\bibitem{seme1}
N. Beisert, C. Kristjansen, J. Plefka, G. W. Semenoff, M. Staudacher, {\sl 
BMN correlators and operator mixing in 
N=4 super Yang-Mills theory}, hep-th/0208178.

\bibitem{freed1}
N.R. Constable, D.Z. Freedman, M. Headrick, S. Minwalla, 
{\sl Operator mixing and the BMN correspondence}, hep-th/0209002.

\bibitem{af}
G. Arutyunov, S. Frolov, {\sl Four graviton scattering amplitude 
from $S^{N}R^8$ 
supersymmetric orbifold sigma model}, Nucl.Phys. {\bf B524 } (1998) 159,
hep-th/9712061.

\bibitem{gg}
R. Gopakumar, D. J. Gross , {\sl Mastering the Master Field }, 
Nucl.Phys. {\bf B451 } (1995) 379, hep-th/9411021.



\end{thebibliography}
\end{document}